# Transits of Venus and Solar diameter measures from ground: method and results from Athens (2004) and Huairou (2012)

C. Sigismondi (1), A. Ayiomamitis (2), X. Wang (3), W. Xie (3), M. Carinci (4), A. Mimmo(4)
(1) ICRA/ICRANet Rome   sigismondi@icra.it
(2) www.perseus.gr   anthony@perseus.gr
(3) Huairou Solar Observing Station and Chinese Academy of Science
(4) Sapienza University of Rome

The variation of the solar diameter in time and in position angle has implications in astrophysics and in general relativity, as the long series of studies attest. The Transits of Venus in 2004 and 2012 have been carefully studied because of the rarity of the phenomenon and its historical importance due the AU measure and to the discovery of Venus atmosphere. The characterization of Venus atmosphere and the measure of the solar diameter to the milliarcsecond level of precision have been studied also from satellite images. The results of the solar diameter measurements made with the observations in Athens (2004) and at the Huairou Solar Observing Station in China (2012) are presented. The topic of the oblateness of the Sun at sunset and its intrinsic value is drafted to introduce the general public to the relativistic relevance of measuring the solar figure, in the occasion of the International Year of Light 2015.

Keywords Astrometry, Solar Diameter, Transits of Venus

**Tthe interest on the variations of the solar radius in General Relativity and classical physics**
Besides the long debate about the sphericity of the Sun and the verifications of the theory of General Relativity and alternative theories, already presented in this meeting to this same parallel section (Sigismondi, et al. 2015), there is a mere thermodynamical fact related to the compression and expansion of the outer shell of the Sun, eventually measurable as shrinking or enlargement of the solar radius. This argument has implications in classical astrophysics and to climatic (and economic) history of the World, as the little ice age (1645-1715 Maunder minimu) and the global warmings (ancient climatic optimum and current) witness.

**Historical transits of Venus**
The first transit of Venus observed was on December 4[th], 1639 by Jermey Horrocks; its observation was done for two main reasons, the verification of the prediction of the transit itself after the observation of the relative trajectoryof the planet with respect to the Sun, and the evaluation of the distance Earth-Sun.

In effect this first observation occurred on Sunday and the only one observer was on religious service in England, and he could observe the phenomenon only two times in the afternoon, moreover the phenomenon itself started one hour before the sunset.

The results of these first observations were already interesting:
by projection it was possible to measure the relative dimensions of Venus and the Sun, by which Horrocks evaluated the distance of Venus; the method was empyrical but the bounty of the measurement of the diameter of Venus against the photosphere was good.

The observation itself of the phenomenon set a cornerstone in the ephemerides of Venus.

The evaluation of the angular diameter of Venus in the inferior conjunction during the transit of the Sun is a situation more favorable with respect to other geometries with the planet over a dark background. Horrocks measured it as 76"±4" (F. Jackson, in Biography of Astronomers)[1] while the real value was 63.14", but this deduction was also related on the accuracy on the solar diameter measurement. Horrocks evaluated the diameter of Venus as 1/30 of the solar diameter. This proportion with the correct value of the solar diameter would give a diameter of 64.98", very well in

---
[1] https://books.google.it/books?id=t-BF1CHkc50C&pg=PA528&lpg=PA528&dq=venus+diameter+horrocks&source=bl&ots=mg59lYMyC8&sig=41DpnnvvMWyEGvb4SZXcc0kZpqQ&hl=it&sa=X&ei=UwiQVdj6PIGbsgHLiYjADg&ved=0CEoQ6AEwBQ#v=onepage&q=venus%20diameter%20horrocks&f=true

agreement with the modern ephemerides of 63.14".

The "irradiation effect" due to the diffraction of the objective and the luminosity of the solar limb over the surrounding background could affect these early measurements. The projected disk of the Sun was 6 inches (15.24cm, A. Chapman 1990)[2] large and presumably the lens was 2 or 3 cm wide. So the diffraction was about 5" and the irradiation larger.

We know the efforts of Francesco Bianchini to observe Venus at or before sunset from the Palatine Hill in Rome with the longest focal telescopes available at his times, the ones built by Giuseppe Campani in the atelier of Montecitorio, 50 meters of focal length. The effects of the chromatic distortions of these single lens objectives were reduced with the long focal lengths, working with paraxial rays, but the spherical aberrations (as in the lens of Cassini meridian line in Paris Observatory) and the great brightness of Venus and the heavy atmospheric turbulence near the horizon made these observations complicate.

Bianchini could achieve a surpringly high precision, within 0.3 arcseconds, published in his book *Hesperi et Phosphori* of 1728, the year before his death.[3]

During the transit of 1761 among other successes, related to the measurement of the Astronomical Unit, Mikhail Lomonossov first found the evidences of the Venus' atmosphere. Its ability to observe and the perfectioned optics of the telescopes, particularly clean, allowed this significant achievement. About the other transits, always in couples spaced each 108 years, 1769 (again on the astronomical unit and the black drop problem), 1874 and 1882 (after the advent of the photography)

In a paper of Sigismondi (2012) there are some more details on these transits, while in this one we want to focus on diameters and irradiation effects.

The last two transits in the digital era are the first to be exploited to measure the solar diamter overcoming the black drop effect, due to the interplay between limb darkening function of the Sun and the diffraction pattern of the telescope. The extrapolation of the time in which the chord cast by the disk of Venus on the solar limb was zero, and so recovering the ingress and egress time, give the calculations of the duration of the transit which is propotional to the diameter of the Sun.

**The method of recovering the solar diameter from planetary transits**

The idea to use a planetary transit to recover the diameter of the Sun has been presented by Irwin Shapiro in 1980 to verify the affirmation that the solar diameter shrinked since 1715 to 1979 according to total solar eclipse observations (SOLE, 1979). The proportion between the duration of the transit and the diameter was used, according to the first order solution of the equation of the length of the measured chord. Nevertheless the velocity of Venus across the solar disk was not constant during the transit, and it was observed in Rome with the 120/1000 Konus refractor and a K12 eyepiece with a haircross during that transit by Sigismondi (2004). So the proportion $\Delta T/T=(T_{obs}-T_{calc})/T_{calc}=\Delta R/R$ valid to the first order, has to be corrected because of the different angular velocities at the ingress and at the egress. Here we do these calculations for the transits of 2004 and 2012.

**The transit of Venus in 2004: data in Halpha from Athens and in visible light from Rome**

Ingress and egress lasted respectively 19m25s and 19m06s; being the Venus' diameter of 57.74", the radial velocity of Venus at ingress is 0.0495"/s: it is the relative angular velocity of Venus with respect to the Sun; it has been calculated from the ephemerides along a solar radius.

The visual observations of Sigismondi (2004) from Rome helped us to understand this phenomenon and the difference of velocity between ingress and egress, due to the vectorial composition of Venus and Earth's orbital velocities and Earth's rotation velocity, different from early morning (directed toward the Sun) to noon (maximum tangential value).

The radial velocity of Venus at egress is 0.0508"/s.

The solar diameter according to the standard value at the day of the transit 08 June 2004 is 1891.66", with a radius of 945.83"; the correction of the solar radius for the ingress, according to the ephemerides with the standard Sun, is 372 mas.
The correction of the solar radius for the egress is 390 mas.
The effect of the atmospheric turbulence are translated into statistical errors of the determination of the instants of t1 and t3 are respectively: 8.07s and 2.63s, which in milliarcsec are
132 mas at t3 (around local noon, with Sun and Venus at 70° above the horizon)
400 mas at t1 one hour after sunrise, with Sun and Venus at 25° above the horizon, strongly affected by atmospheric turbulence.
Since each fit of the chord's length, draft by the disk of Venus over the limb of the Sun, has been made with 21 datapoints, the average seeing ρ can be obtained from the statistical errors ρ=√21·σ obtaining a seeing of 0.6" at local noon, 70°, and 1.8" in the early morning at 25° of altitude.
The average value of the corrections at the ingress and at the egress is 381 mas, with the maximum error of 9 mas.
The normalized value of the corrections to the solar radius, to the distance of 1AU, is 387±9 mas.

---
[2] http://articles.adsabs.harvard.edu/cgi-bin/nph-iarticle_query?1990QJRAS..31..333C&defaultprint=YES&filetype=.pdf
[3] Sigismondi, C., La lente di G. Campani alla meridiana di J. Cassini a Parigi e quelle usate da F. Bianchini nelle osservazioni di Venere a Roma, Atti convegno SIA 2012 http://www.ibs.it/code/9788882924867/atti-del-12deg;.html
http://www.brera.inaf.it/archeo/12-convegno/Programma2012finale.pdf

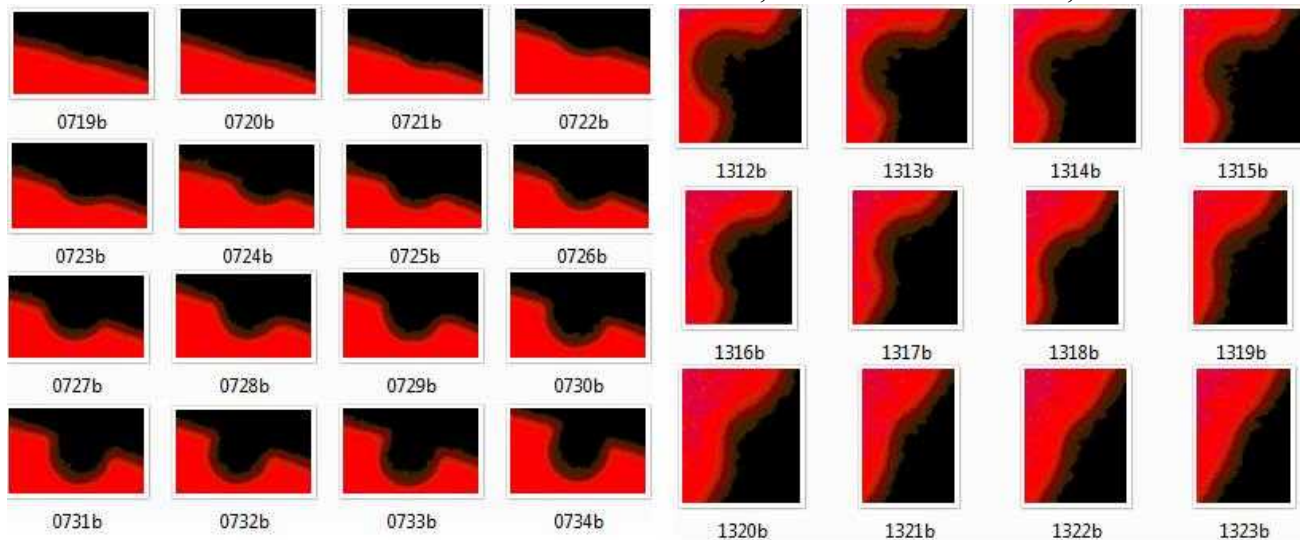

Fig. 1 Phases of the Ingress (left) and of the Egress of the 2004 transit in Hα, times in UT.

**Results from satellite images**

The analysis of SOHO data on the transit of Mercury of 2003 and 2006 has been done by Marcelo Emilio and his collaborators in two times 2004[4] and 2012,[5] with a substancial recorrection of the first value of the solar radius at 676.78 nm measured with the SOHO/MDI instrument. The solar radius at 676.78 nm normalized at 1 AU has been found at 960.12". The transit of Venus has not been observed b y SOHO since it was not on the line of sight of the transit (it is in the Earth-Sun L1 Lagrangian point at 1.5 million Km from the Earth), but SDO has observed it. Again M. Emilio has studied this 2012 Venus transit and found the radius of the Sun at 617.3 nm being 959.90".[6]

**Comparison between satellite data and our value of 2004 solar radius in H-alpha**

Interpolating the results of M. Emilio of 2012 and 2015 we have the following results, referred to 959.63" as the standard solar radius (Auwers, 1891).[7]

| λ [nm] | ΔR [mas] milliarcsec | Source |
|---|---|---|
| 617.3 | 270±6 | M. Emilio et al. (2015) |
| 676.78 | 490±9 M. Emilio et al. (2012)+ | J. R. Kuhn, et al. (2004) |
| 656.281 (Hα) | 414±11 | Extrapolated from 617 and 676 |
| 656.281 (Hα) | 387±9 | This paper for 2004 transit (i+e) |
| 551±88 (V) | 280±159 | This paper for 2012 egress only |

The value measured in the Hα line is in good agreement with the value extrapolated from the two measures in two wavelengths of the continuum of higher 676 and lower 617 lambda. Meftah et al., 2014 found 959.78"±0.19" with the ground-based SODISM2: ΔR=+160±190 in agreement with us.

**The Observing campaign in Huairou and the search for Venus aureola in 2012**

The analysis of the data from Huairou 2012 is progressing:

the radial velocity of Venus at ingress is $5.54 \cdot 10^{-2}$"/s and at the egress is $5.63 \cdot 10^{-2}$"/s, 1.6% larger. The solar diameter according to the standard value at the day of the transit 06 June 2012 is 1891.66" the correction of the solar radius for the ingress, according to the ephemerides with the standard Sun, is undergoing a further analysis because of the poor signal-to-noise, due to the presence of clouds over a diffuse haze at sunrise, while the correction of the solar radius for the egress is +4.9±2.8 s corresponding to ΔR=+280±159 mas at 1AU, in agreement with Meftah et al. 2014 with a narrower errorbar. The effect of the atmospheric turbulence from the statistical errors of the determination of the egress instants is corresponding to a seeing of $0.157 \cdot \sqrt{8}=0.44$", due to the position on the top of the tower over the lake of Huairou, a chinese replica of Big Bear Solar Observatory.[8] This analysis has been made using as fitting function the first order approximation (the Sun is a line) of the analytical formula for the length of the chord drawn by Venus on the solar limb. The chord $C=\sqrt{[8 \cdot R \cdot v \cdot (t_o \pm t)]}$ with the time t after/before the external contacts, R is the radius of Venus in pixels (23.5 fixed value obtained from the images of the transit with the full disk of Venus) and the fixed velocity v=0.451px/s for the egress (0.444, ingress) using the ephemerides.

The observational campaign of 2012 was part of an international effort guided at the Observatory of the Côte d'Azur in France by P. Tanga[9] to measure the aureola of Venus; first evidence of the venusian atmosphere for M. Lomonossov in 1761. The aim was to characterize the parameters of the upper atmosphere of Venus studying its refracting properties,[10] and the solar diameter's measure.

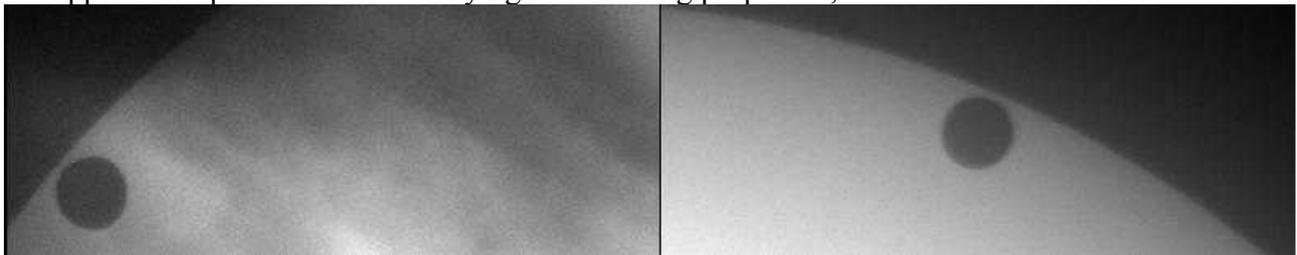

Fig. 2 The haze at the ingress (left) reduced the signal to noise, and the time of the contact is to be refined, while the egress was under ideal conditions and allowed us to perform a good measure.

**Conclusions: the results from Athens and Huairou**

Both the measurements of 2004 and 2012 transits achieved an accuracy of single determinations of ΔR=±0.15", and it reduces to ΔR=±9 mas using both contacts. These are the best measurements achievable from the ground. Next transit of Mercury of May 9, 2016 it is a good occasion to implement these formulae and this method to recover a measurement of the solar diameter to the nearest milliarcsec of accuracy, at least at the contact visible at noon time. The opportunity given by the next transit is exploitable even observing only the ingress: the ephemerides of the positions of the planet and of the Sun are nowadays very reliable and can be exploited once again to correct the standard value of the solar diameter. This approach will be perfectioned by using the first order formula $C=2\sqrt{v \cdot (t_o \pm t)/a \cdot (1-1/(1+2a \cdot r))}$, with 1/2a=R_Sun: a more realistic Venus-(round) Sun chord

---

8  http://www.bbso.njit.edu/  Huairou is on http://sun.bao.ac.cn/ a 12" SC telescope has been used for the occasion.

9  http://adsabs.harvard.edu/abs/2012DPS....4450807T Tanga, P. et al. 2012.

10 http://adsabs.harvard.edu/abs/2013IAUS..294..485X Xie, W., et al. 2013.